\documentclass[aps,pra,reprint,superscriptaddress]{revtex4-1}
\pdfoutput=1
\usepackage{graphicx}

\usepackage{amsmath}
\usepackage{braket}
\usepackage{textcomp}
\usepackage[utf8]{inputenc}

\newcommand{\msx}{\textnormal{MS}_x}
\newcommand{\ex}{\textnormal{MS}_x}

\newcommand{\cnot}{\textnormal{CNOT}}

\newcommand{\coll}{C}
\newcommand{\id}{\mathbf{1}}
\newcommand{\prodx}[1]{\tilde{#1}}

\begin{document}

\title{Compiling quantum algorithms for architectures with multi-qubit gates}
\author{Esteban A. Martinez}
\author{Thomas Monz}
\author{Daniel Nigg}
\author{Philipp Schindler}
\affiliation{Institut f\"ur Experimentalphysik, Universit\"at Innsbruck, Technikerstraße 25/4, 6020 Innsbruck, Austria}
\author{Rainer Blatt}
\affiliation{Institut f\"ur Experimentalphysik, Universit\"at Innsbruck, Technikerstraße 25/4, 6020 Innsbruck, Austria}
\affiliation{Institut f\"ur Quantenoptik und Quanteninformation, \"Osterreichische Akademie der Wissenschaften, Technikerstraße 21a, 6020 Innsbruck, Austria}

\begin{abstract}  
In recent years, small-scale quantum information processors have
been realized in multiple physical architectures. These systems
provide a universal set of gates that allow one to implement any
given unitary operation. The decomposition of a particular
algorithm into a sequence of these available gates is not
unique. Thus, the fidelity of the implementation of an algorithm
can be increased by choosing an optimized decomposition into
available gates. Here, we present a method to find such a
decomposition, taking as an example a small-scale ion trap quantum
information processor. We demonstrate a numerical optimization
protocol that minimizes the number of required multi-qubit entangling gates
by design. Furthermore, we adapt the method for
state preparation, and quantum algorithms including in-sequence
measurements.
\end{abstract}

\maketitle

\tableofcontents

\section{Introduction}
\label{sec:introduction}
Quantum technologies open new possibilities that are inaccessible with current classical devices, ranging from cryptography \cite{Ekert1996,Gisin2002} to efficient simulation of physical systems \cite{Cirac2012,Bloch2012,Blatt2012}. To utilize the full computational power of quantum systems, one needs a \emph{universal quantum computer}: a device able to implement arbitrary unitary operations, or at least to approximate them to arbitrary accuracy. However, in any specific physical system, only a certain set of operations is readily available. Therefore, it is necessary to decompose the desired unitary operation as a sequence of these experimentally available gates. An available set of gates is known as \emph{universal} if it is possible to find such a decomposition for an arbitrary unitary quantum operation acting on the qubit register.

A canonical universal set of gates consists of two-qubit CNOT gates and arbitrary single qubit rotations. There exist deterministic algorithms that provide near-optimal decompositions of unitaries in terms of these gates \cite{Nielsen2004}. However, the set of gates that yields the highest fidelities depends on the particular experimental implementation. In particular, two-qubit CNOT gates may not be the most efficient to implement. Architectures like trapped ions \cite{Schindler2013,Harty2014} or atom lattices \cite{Xia2015} include in their toolboxes high-fidelity multi-qubit gates that act on the entire qubit register (see section \ref{sec:toolbox}). Implementing two-qubit gates in terms of these requires refocusing \cite{Vandersypen2005} or decoupling \cite{Schindler2013} techniques, and thus increases the overhead. Therefore it is desirable to find a direct decomposition of the target unitary into the available operations. In general, the number of multi-qubit gates needs to be minimized, since these are more prone to errors than local gates.

Compiling unitaries using multi-qubit gates that act on the whole qubit register is more challenging than using two-qubit gates. Even if a sequence implements correctly a unitary for $N$ qubits, it might not work for $N+1$ qubits, since additional ``spectator'' qubits will also be affected by the sequence instead of being left unchanged~\cite{Nebendahl2009}. Therefore, one has to define a qubit register of interest where the unitary will be compiled, and the experimental implementation of the resulting sequence has to be limited to this subregister, as explained in Section~\ref{sec:toolbox}. Moreover, the existing analytical methods for finding decompositions of unitaries in terms of two-qubit gates (see for instance Ref.~\cite{Khaneja2001}) do not seem to apply to multi-qubit gates. Therefore, in this work we employ an approach based on numerical optimization.

A similar algorithm for finding multi-qubit gate decompositions has been studied in Ref.~\cite{Nebendahl2009}, where optimal control techniques are used to find a pulse sequence for a given target unitary operation. The procedure described there starts with long sequences and then removes pulses, if possible. This often results in sequences with more entangling operations than actually required. In this work we present an algorithm designed to produce decompositions with a minimal number of entangling gates. In addition, we introduce a deterministic algorithm for finding decompositions of local unitaries. We also extend the algorithm to operations required for state preparation or measurement, which are particular cases of more general operations known as \emph{isometries}~\cite{Knill1995,Iten2016}.

The paper is organized as follows: in Section~\ref{sec:toolbox} we describe precisely which gates we will consider as part of our experimentally available toolbox, and review some architectures for quantum information processing to which the methods described in this work can be applied. In Section~\ref{sec:local} we show an analytic algorithm to compile local unitaries, which can be used to find efficient implementations of state and process tomographies. Finally, in Section~\ref{sec:algorithm} we describe and analyze an algorithm to compile fully general unitaries which relies on numerical optimization.

\subsection{Experimental toolbox}
\label{sec:toolbox}

Several quantum information processing experiments based on atomic and molecular systems have similar toolsets of quantum operations at their disposal. Often, it is convenient to apply collective rotations on an entire qubit register. These collective (yet local) gates, combined with addressed operations (typically rotations around the Z axis) allow one to implement arbitrary local unitaries, as we show in Section \ref{sec:local}. Together with suitable multi-qubit operations, arbitrary quantum unitaries can be implemented. In this work we consider the following set of gates:

\begin{itemize}

\item Collective rotations of the whole qubit register about any axis on the equator of the Bloch sphere $\coll(\theta, \phi)$. Here $\theta$ is the rotation angle and $\phi$ is the phase, so that:
\begin{equation}
\coll(\theta, \phi) = e^{-i \theta (S_x \cos{\phi} + S_y \sin{\phi}) / 2},
\end{equation}
where $S_{x, y} = \sigma_1^{x, y} + \cdots + \sigma_N^{x, y}$ are the total spin projections on the \emph{x} or \emph{y} axes, and $\sigma_j^{x, y, z}$ are the respective Pauli operators corresponding to qubit $j$. For the sake of brevity we also define rotations around the X and Y axes as:
\begin{align}
X(\theta) &= C(\theta, 0),\\
Y(\theta) &= C(\theta, \pi / 2).
\end{align}

\item Single qubit rotations around the $Z$ axis $Z_n(\theta)$, where $\theta$ is the rotation angle, and $n$ is the qubit index:
\begin{equation}
\rm{Z}_n(\theta) = e^{-i \theta \sigma_n^{z} / 2},
\end{equation}
with $\sigma_n^{z}$ being the Pauli Z operator applied to the \emph{n}-th ion.

\item Entangling M{\o}lmer-S{\o}rensen (MS) gates \cite{Sorensen2000}, with arbitrary rotation angle and phase $\operatorname{MS}_{\phi}(\theta)$. Here $\theta$ is the rotation angle and $\phi$ is the phase of the gate, resulting in:
\begin{equation}
\operatorname{MS}_{\phi}(\theta) = e^{-i \theta (S_x \cos{\phi} + S_y \sin{\phi})^2 / 4},
\end{equation}
where $S_{x, y} = \sigma_1^{x, y} + \cdots + \sigma_N^{x, y}$ are the total spin projections on the \emph{x} or \emph{y} axes, as before. For $\phi = 0$ or $\phi = \pi / 2$ we obtain gates that act around the $X$ or $Y$ axes, which we will denote:
\begin{equation}
  \rm{MS}_{x,y}(\theta) = e^{-i \theta S_{x,y}^2 / 4}.
\end{equation}
As mentioned before, it is desirable to be able to restrict the action of the MS gate to a particular qubit subset. This can be done experimentally by spectroscopically decoupling the rest of the qubits from the computation~\cite{Schindler2013}, or by addressing the MS gate only on the relevant subset of the qubits~\cite{Debnath2016}.

\end{itemize}

This set of gates, or equivalent ones, are available in several trapped-ion experiments~\cite{Gaebler2012,Schindler2013,Monroe2014}.
A similar toolbox of operations is available for architectures based on trapped-ion hyperfine qubits. For example, Ref.~\cite{Harty2014} describes high-fidelity microwave gates applied to a single hyperfine $^{43}$Ca$^+$ qubit. In a multi-qubit system, these gates would drive collective rotations like the ones described above. In addition, Ref.~\cite{Ballance2015} describes a Raman-driven $\sigma^z \otimes \sigma^z$ phase gate on two qubits, which applied to a many-qubit register would act analogously to the MS gate already described.

Recently, an implementation of high-fidelity gates in a 2D array of neutral atom qubits was reported \cite{Xia2015}. The toolbox described there consists of global microwave-driven gates and single-site Stark shifts on the atoms, which are completely equivalent to the local operations described before for the trapped ion architecture. A multi-qubit CNOT gate, equivalent to the MS gate, could also be implemented by means of long-range Rydberg blockade interactions \cite{Isenhower2011}.

\section{Compilation of local unitaries}
\label{sec:local}

\begin{figure}[ht]
\centering
\includegraphics[width=\columnwidth]{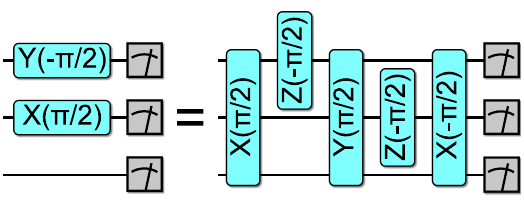}
\caption{Pulse sequence to perform a projective measurement on the $\{\text{X}, \text{Y}, \text{Z}\}$ bases for qubits $\{1, 2, 3\}$, respectively.}
\label{fig:local}
\end{figure}

\emph{Local} unitaries can be written as a product of single-qubit unitaries. In this section we show a fully deterministic algorithm that produces decompositions of any local unitary as a sequence of collective equatorial rotations and addressed Z rotations, as described in section \ref{sec:toolbox}. The decompositions presented here are optimal in the number of pulses. These techniques are particularly useful for the implementation of state and process tomographies, as exemplified in figure \ref{fig:local}, since both require only local operations at the beginning and end of the algorithm.


Let us consider a register of $N$ qubits, and a local unitary $U = U_1 \otimes U_2 \otimes \dotsm \otimes U_N$ to be applied to them, where $U_i$ is the action of the unitary on the $i$-th qubit. If the same operation has to be applied to more than one qubit ($U_i = U_j$), we can replace both with a single instance of the operation, and then apply the same addressed rotations on all qubits subject to the same operation $U_i$. Therefore, we only have to consider the case where every $U_i$ is unique.

In order to apply a general local unitary to each qubit we need to have at least three degrees of freedom per qubit \cite{Nielsen2004}, so the decomposition must have at least $3N$ free parameters. During the sequence at least $N - 1$ of the qubits must eventually be addressed, since a different unitary has to be applied to each qubit. Therefore, a sequence of addressed operations of the form $Z_1(\theta_1), Z_2(\theta_2), \dotsc Z_{N-1}(\theta_{N-1})$ must be included in the decomposition. These provide $N - 1$ parameters, so $2N + 1$ more degrees of freedom are required. The most economic way to provide these is by means of collective gates $\coll(\theta_i, \phi_i)$, which have two degrees of freedom each, so the shortest sequence possible must include at least $N$ global operations, for a total of $3N - 1$ free parameters. One additional degree of freedom remains, so we must add a last gate. This can be either an addressed operation on qubit $N$ or a collective gate. If we add an addressed gate $Z_N$, we obtain a sequence of the form:
\begin{equation}
\label{eq:complete-z}
U = Z_N \coll_N Z_{N-1} \coll_{N-1} \dotsm Z_2 \coll_2 Z_1 \coll_1,
\end{equation}
where $C_i = C(\theta_i, \phi_i)$ and $Z_i = Z_i(\theta_i)$ are collective and single-qubit rotations respectively, as explained in Section \ref{sec:toolbox}. Such a sequence is useful for compiling local unitaries up to arbitrary phases, as explained in Appendices \ref{sec:up-to-collective} and \ref{sec:up-to-independent}. The second alternative is to add a collective rotation $C'_N$:
\begin{equation}
\label{eq:complete-g}
U = \coll_N' \coll_N Z_{N-1} \coll_{N-1} \dotsm Z_2 \coll_2 Z_1 \coll_1,
\end{equation}
which is the type of sequence we consider in this section.

For particular unitaries, some of the $C_i$ and $Z_i$ in Eq.~(\ref{eq:complete-g}) may actually be the identity, in which case the sequence is simpler. Since the decomposition depends on the ordering of the qubits, by reordering them a simpler sequence might be obtained. For small numbers of qubits, one can compile the unitary for every possible permutation, although this becomes inefficient for large numbers of qubits. However, let us remember that, for the purposes of the compilation, the qubits are grouped together according to which of them experience the same single-qubit unitary $U_i$. For an application such as state tomography, there are only three possible unitaries to be applied to each qubit in the register (shown in Figure~\ref{fig:local}), since one only wants to perform a measurement in one of three different bases. Therefore, effectively we only need to consider three qubits, in which case trying out all the permutations is perfectly feasible.

We will describe now how to compile a generic local unitary $U = U_1 \otimes U_2 \otimes \dotsm \otimes U_N$ exactly, using a decomposition of the form (\ref{eq:complete-g}). Let us first note that the unitaries in Eq.~(\ref{eq:complete-g}) act on the $N$-qubit Hilbert space, which is the tensor product of the single-qubit Hilbert spaces. For the sake of simplifying the notation, we will now refer to these unitaries as $\tilde{C}_i$, $\tilde{Z}_i$, and will reuse the notations $C_i$ and $Z_i$ for their action on the single qubits, so that:
\begin{align}
\tilde{C}_i &= C_i \otimes C_i \otimes \dotsm \otimes C_i, \\
\tilde{Z}_i &= \id \otimes \id \otimes \dotsm \otimes Z_i \otimes \cdots \otimes \id, \nonumber
\end{align}
where $\id$ is the $2 \times 2$ identity matrix, and $Z_i$ appears at the $i$-th place (since it only addresses the $i$-th qubit).

In terms of these single-qubit unitaries, factoring Eq.~(\ref{eq:complete-g}) for each qubit we obtain $N$ equations:
\begin{align}
U_1 &= \coll_N' \coll_N \dotsm \coll_2 Z_1 \coll_1,\\
U_2 &= \coll_N' \coll_N \dotsm  Z_2 \coll_2 \coll_1, \nonumber\\
&\ \ \vdots \nonumber\\
U_N &= \coll_N' \coll_N \cdots \coll_2 \coll_1. \nonumber
\end{align}

From the last equation we can determine $\coll_N' \coll_N$:
\begin{equation}
  \label{eq:last-u}
  \coll_N' \coll_N = U_N  \coll_1^{-1} \coll_2^{-1} \dotsm \coll_{N-1}^{-1},
\end{equation}
and eliminating this factor from the remaining equations we obtain:
\begin{align}
  \label{eq:system-eliminated}
U_N^{-1} U_1 &= \coll_1^{-1} Z_1 \coll_1,\\
U_N^{-1} U_2 &= \coll_1^{-1} \coll_2^{-1} Z_2 \coll_2 \coll_1,\nonumber\\
&\ \ \vdots \nonumber\\
U_N^{-1} U_{N-1} &= \coll_1^{-1} \coll_2^{-1} \dotsm \coll_{N-1}^{-1} Z_{N-1} \coll_{N-1} \dotsm \coll_2 \coll_1.\nonumber
\end{align}

We solve each equation in (\ref{eq:system-eliminated}) consecutively. To solve the first equation in (\ref{eq:system-eliminated}), let us notice that its left-hand side is a known unitary, which can be written as:
\begin{equation}
  \label{eq:generator}
  U_N^{-1} U_1 = e^{-i \alpha_1 u_1 / 2},
\end{equation}
where $\alpha_1$ is the angle of the rotation and $u_1$ its generator. The right-hand side is simply a rotation around Z and a change of basis. Therefore, the rotation angle of $Z_1$ must be equal to $\alpha_1$, and the change of basis must be such that:
\begin{equation}
  \label{eq:basis-change}
  u_1 = \coll_1^{-1} \sigma_z \coll_1.
\end{equation}
We show in Appendix \ref{sec:ap-exact} how to find the generator and angle of the collective rotation $\coll_1$.

Having determined $\coll_1$, we can write the second equation in (\ref{eq:system-eliminated}) as:
\begin{equation}
  \coll_1 U_N^{-1} U_2 \coll_1^{-1} = \coll_2^{-1} Z_2 \coll_2.
\end{equation}
As before, the left-hand side of this equation is a known unitary, and the right-hand side consists of a rotation around Z and a change of basis, so the rotation angle $\theta_2$ and generator of the change of basis $C_2$ can be found as for the previous equation. This procedure can be repeated until all of the $\coll_k$ and $Z_k$ with $k \leq N - 1$ are determined. The last collective operations $C_N$ and $C_N'$ can be determined from equation (\ref{eq:last-u}). For this we need to decompose an arbitrary unitary into a product of two equatorial rotations; this can be done as explained in appendix \ref{sec:ap-two-coll}.

We have shown so far how to compile a local unitary exactly. However, in certain cases the constraints on the target unitary are weaker, so that it can be implemented with a simpler sequence. For instance, a unitary that is followed by global gates whose phase can be freely adjusted must only be specified up to a collective Z rotation afterwards, since this rotation can be absorbed into the phase. This removes one free parameter from the sequence, thus simplifying its implementation. The details of this procedure are presented in appendix \ref{sec:up-to-collective}. Another case of interest is when the target unitary is specified up to arbitrary independent Z rotations afterwards, for instance when the unitary is followed by a projective measurement on the Z basis. This is particularly useful for tomographic measurements; details are shown in Appendix~\ref{sec:up-to-independent}.

\section{Compilation of general unitaries}
\label{sec:algorithm}
\begin{figure}[ht]
\centering
\includegraphics[width=\columnwidth]{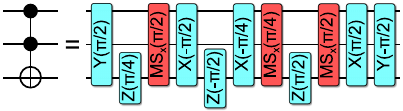}
\caption{Decomposition of a Toffoli gate into a pulse sequence of collective equatorial rotations, addressed Z rotations and entangling M{\o}lmer-S{\o}rensen (MS) gates.}
\label{fig:toffoli}
\end{figure}

In section \ref{sec:local} we studied how to compile local unitaries in terms of collective and addressed rotations. However, a universal quantum computer also requires entangling unitaries, which must be compiled into the experimentally available local and entangling gates. For example, in figure \ref{fig:toffoli} we show a decomposition of a Toffoli gate into a sequence of local and entangling gates applied consecutively. In this section, we present an algorithm to find such decompositions for arbitrary unitaries.

We seek decompositions directly in terms of multi-qubit entangling gates, since these are often more efficient than decompositions in terms of two-qubit gates. For example, a Toffoli gate can be implemented using only 3 M{\o}lmer-S{\o}rensen (MS) gates \cite{Nebendahl2009}, while 6 CNOT gates are needed to implement it \cite{Shende2009}), and a Fredkin gate can be implemented using 4 MS gates \cite{Monz2016}, while the least number of two-qubit gates required is 5 \cite{Yu2013}. As described in section \ref{sec:toolbox}, many equivalent types of entangling gates are experimentally available. We will consider MS gates, but the methods shown here are applicable to any entangling gate that forms a universal set together with local operations.

\subsection{Compilation in layers}
\label{sec:compilation-in-layers}
\begin{figure}[ht]
\centering
\includegraphics[width=\columnwidth]{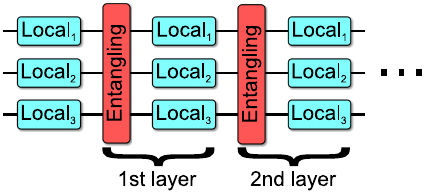}
\caption{Sequence with layers of local and entangling gates applied consecutively.}
\label{fig:layers}
\end{figure}

In many quantum information processing experiments the most costly operations in terms of fidelity are entangling gates. Therefore, when trying to compile a unitary we seek to minimize the number of those. A straightforward way to do this is to use pulse sequences where layers of local unitaries and entangling gates are applied consecutively, as shown in figure \ref{fig:layers}.

Any unitary can be decomposed in terms of single-qubit gates and two-qubit CNOT gates \cite{divincenzo}:
\begin{equation}
\label{eq:divincenzo}
U = L_M \ \cnot_{M} \ \dotsm \ L_{1} \ \cnot_1 \ L_0,
\end{equation}
where $L_i$ denotes an arbitrary local unitary on the whole qubit register and $\cnot_i$ denotes a gate between some two qubits. A two-qubit $\cnot$ gate can be implemented in an arbitrary $N$-qubit register as a sequence of local unitaries and $\msx(\pi / 8)$ gates \cite{Nebendahl2009}. Therefore, the following decomposition is always possible:
\begin{equation}
\label{eq:divincenzo-ms}
U = L_M \ \ex(\pi / 8) \ \dotsm \ L_{1} \ \ex(\pi / 8) \ L_0.
\end{equation}
However, some of the local unitaries $L_i$ in a decomposition of the form (\ref{eq:divincenzo-ms}) may actually be identity, so after removing them the resulting sequence has the following structure:
\begin{equation}
\label{eq:seq}
U = L_{M} \ \ex(k_M \pi / 8) \ \dotsm \ L_{1} \ \ex(k_1 \pi / 8) \ L_0,
\end{equation}
where the $k_i$ are integers, and the number $M$ of entangling gates is the same or less than in Eq.~(\ref{eq:divincenzo-ms}). It is enough to consider $0 \leq k \leq 7$, since $\ex(\pi)$ is either the identity for an odd number of qubits, or a $\pi$ rotation around $X$ for an even number of qubits.

We now seek to further simplify sequence (\ref{eq:seq}). Every single-qubit unitary $U_i$ on qubit $i$ can be written as a composition of rotations around two different fixed axes \cite{Nielsen2004}, which means that we can always choose $\alpha_{i1}$, $\alpha_{i2}$ and $\alpha_{i3}$ such that:
\begin{equation}
U_i = X_i(\alpha_{i3}) Z_i(\alpha_{i2}) X_i(\alpha_{i1}).
\end{equation}
Any local unitary $L = \prod_{i = 1}^N U_i$ can therefore be written as:
\begin{equation}
L = \prod_{i = 1}^N X_i(\alpha_{i3}) Z_i(\alpha_{i2}) X_i(\alpha_{i1}),
\end{equation}
where the product goes over the $N$ qubits in the register. Since unitaries acting on different qubits commute, we can write this as:
\begin{align}
L &= \prod_{i = 1}^N X_i(\alpha_{i3}) \prod_{i = 1}^N Z_i(\alpha_{i2}) \prod_{i = 1}^N X_i(\alpha_{i1}) \\
&= \prodx{X'} \prodx{Z} \prodx{X},
\end{align}
where $\prodx{X}$ and $\prodx{Z}$ denote arbitrary products of rotations around the $X$ or $Z$ axes for all qubits. Therefore, the sequence in (\ref{eq:seq}) can be written as:
\begin{align}
U &= \prodx{X}_{M}' \prodx{Z}_{M} \prodx{X}_{M} \ex(k_M \pi / 8) \cdots \times \\
& \quad \times \prodx{X}_1' \prodx{Z}_1 \prodx{X}_1 \ex(k_1 \pi / 8) \prodx{X}_0' \prodx{Z}_0 \prodx{X}_0, \nonumber
\end{align}
and commuting the $X$ rotations with the MS gates we obtain a sequence of the form:
\begin{align}
U &= \prodx{X}_M' \prodx{Z}_M \prodx{X}_M \ex(k_M \pi / 8) \cdots \times \\
& \quad \ \times \prodx{X}_2' \prodx{Z}_2 \prodx{X}_2 \ex(\alpha_2) \prodx{Z}_1 \ex(k_1 \pi / 8) \prodx{X}_0' \prodx{Z}_0 \prodx{X}_0. \nonumber 
\end{align}
Every odd local unitary (except for the last one) is a product of Z rotations on all qubits, and the even local unitaries can be grouped as $L_i = \prodx{X}_i' \prodx{Z}_i \prodx{X}_i$. Moreover, a collective Z rotation can be extracted from each even local unitary $L_i$ and absorbed into the phase of the subsequent MS gates and collective operations to simplify the implementation of $L_i$. Therefore the sequence can be written as:
\begin{align}
\label{eq:sequence-ms-phase}
U &= L_M \operatorname{MS}_{\phi_M}(k_M \pi / 8)  \cdots \times \\
& \quad \times L_2 \operatorname{MS}_{\phi_2}(k_2 \pi / 8) \prodx{Z}_1 \operatorname{MS}_{\phi_1}(k_1 \pi / 8) L_0. \nonumber
\end{align}

We have thus shown that any $N$-qubit unitary $U$ can be decomposed into a sequence of the form shown in (\ref{eq:sequence-ms-phase}). These sequences always have the same structure, which makes it easier to identify patterns if one wants to compile families of unitaries, i.e. unitaries that depend on some tunable parameter.

\subsection{Numerical optimization}
\label{sec:optimization}

We have described a general form of a sequence of local operations and global entangling gates that implements any desired target unitary. It remains to find the actual sequence parameters, that is, the rotation angles and phases of the gates. However, we do not know a priori how many entangling gates will be needed for a given unitary. Therefore we suggest the following algorithm:
\begin{enumerate}
\item Propose a sequence with $M = 0$ entangling gates.

\item Search numerically for the sequence parameters that maximize the fidelity with the target unitary.

\item If the sequence has converged to the desired unitary (i.e. the fidelity equals 1), stop. Otherwise increase $M$ by 1 and go back to step 2.
\end{enumerate}

When performing the numerical optimization in step 2 there might be a number of local optima in addition to the true global optimum, making fully deterministic optimization methods difficult to apply. We apply therefore a \emph{repeated local search}, where an efficient deterministic optimization method is iterated, each time using randomly determined initial conditions. The initial conditions are chosen randomly for every optimization run, as experience has shown us that starting close to previously found local minima does not offer any improvement. The search is finalized whenever the fidelity with the target unitary is above some predefined threshold, or when a maximum number of tries is exceeded. An advantage of this method is that, since each optimization run starts from random initial conditions, these are easy to perform in parallel.

The algorithm chosen for each numerical optimization is the quasi-Newton method of Broyden, Fletcher, Goldfarb, and Shanno (BFGS) \cite{Nocedal1999}. The function to be maximized is the fidelity of the unitary resulting from the pulse sequence with the target unitary. Since we are interested in exact solutions, we reject those solutions whose fidelity (normalized to the maximum value possible) is not equal to $1$ within some tolerance threshold (usually $1\%$). The gradient of the fidelity can be calculated analytically as a function of the sequence parameters, which speeds up the computation as compared to using several evaluations of the fidelity function.

\begin{figure}[ht]
\centering
\includegraphics[width=\columnwidth]{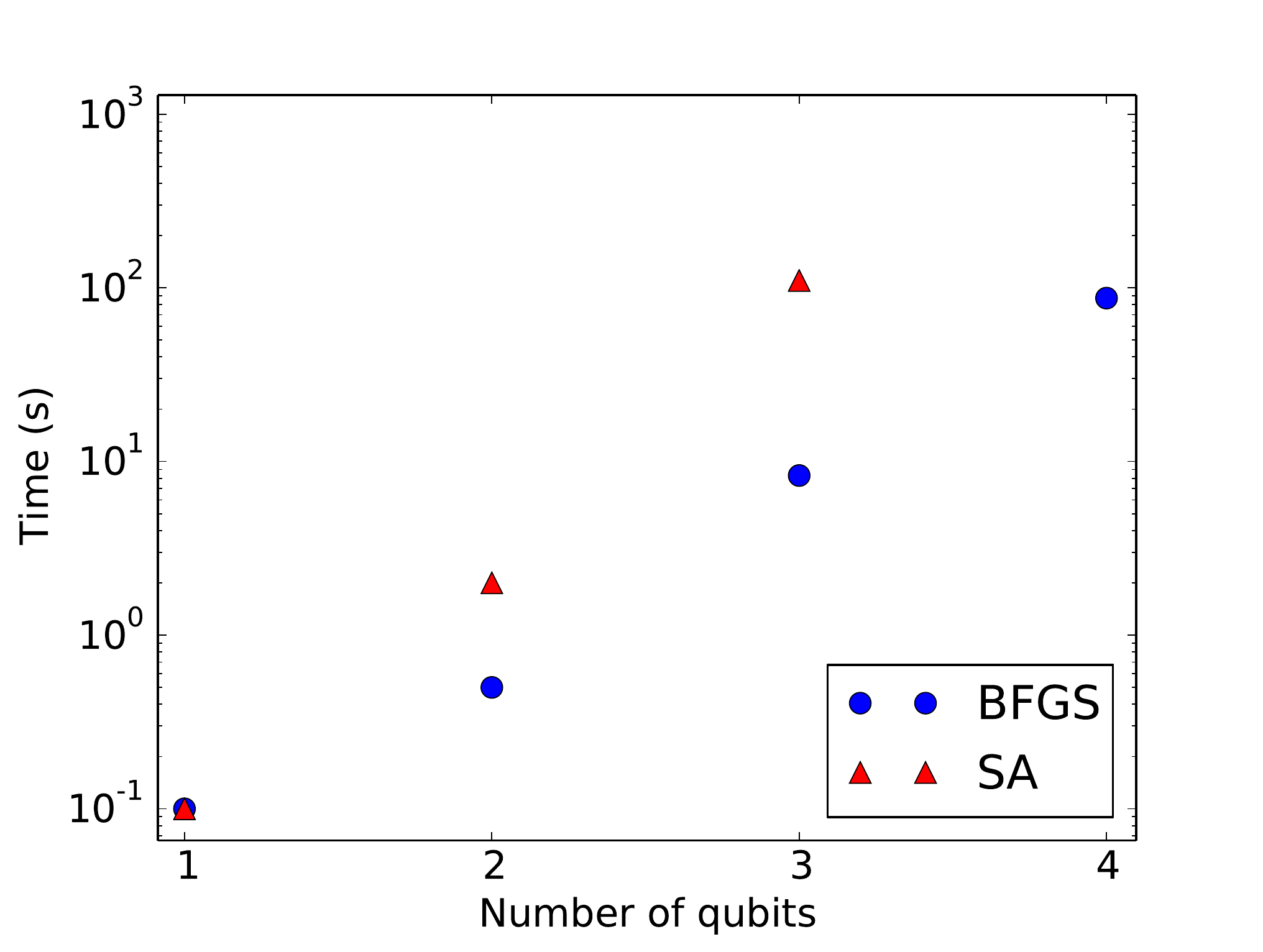}
\caption{Average time required to find the global optimum for 100 unitaries randomly distributed in the Haar measure with the BFGS and simulated annealing methods, using an Intel\textcopyright Core i5-4670s CPU 550 @ 3.10~GHz x 4 (one processing thread per optimization run). No data was obtained for the simulated annealing approach for 4 qubits owing to the excessive time required.}
\label{fig:scaling}
\end{figure}

A previously used approach to this optimization problem was a combination of local gradient descent and simulated annealing (SA) \cite{Nebendahl2009}, which also helps to avoid local maxima. However, this method did not make use of the analytic expression for the fidelity gradient, which speeds up the search. Moreover, its performance depends on the ``topography'' of the optimization space and requires manual tuning of the search parameters to achieve optimal results. We have compared the BFGS and simulated annealing approaches by compiling 100 random unitaries uniformly distributed in the Haar measure as explained in \cite{Mezzadri2007} for different numbers of qubits. In our experience, we find that the BFGS method scales better with the number of qubits than simulated annealing (see Figure~\ref{fig:scaling}). The median number of search repetitions needed to find the global optimum was 1 in all the cases.

The exponential scaling of the optimization problem complexity depends on the number of entangling gates required to compile a given unitary, which is an intrinsic property of the unitary and does not depend on the search algorithm. It is already known that it is not possible to efficiently implement an arbitrary unitary in terms of two-qubit gates \cite{Nielsen2004}; our numerical results suggest a similar result for $N$-qubit gates. In the two-qubit case the compilation always succeeded with 3 entangling gates, and not less (using 200 search repetitions). This was to be expected, since for two qubits an MS gate is equivalent to a $\operatorname{CNOT}$ gate, and it is known that 3 CNOT gates are enough (and in general necessary) to implement an arbitrary two-qubit unitary \cite{Vatan2004,Hanneke2009}. In the three-qubit case, the optimization always succeeded with 8 entangling gates, and never with fewer (also using 200 repetitions). For 4 qubits, the optimization always succeeded for 25 entangling gates, and succeeded only 4\% of the time with 24 entangling gates. However, we did only 4 optimization repeats in the four-qubit case, owing to the increased time it takes for these to converge. Therefore, it might be the case that given enough optimization runs, more unitaries would have been compiled with only 24 gates. We are not aware of any result in the literature concerning the number of $N$-qubit global entangling gates required for implementing a general $N$-qubit unitary for more than $N = 2$ qubits. From our numerical results, we conjecture that any three-qubit unitary can be implemented using at most 8 MS gates, and any four-qubit unitary using at most 24 or 25 MS gates.

\begin{figure}[ht]
\centering
\includegraphics[width=\columnwidth]{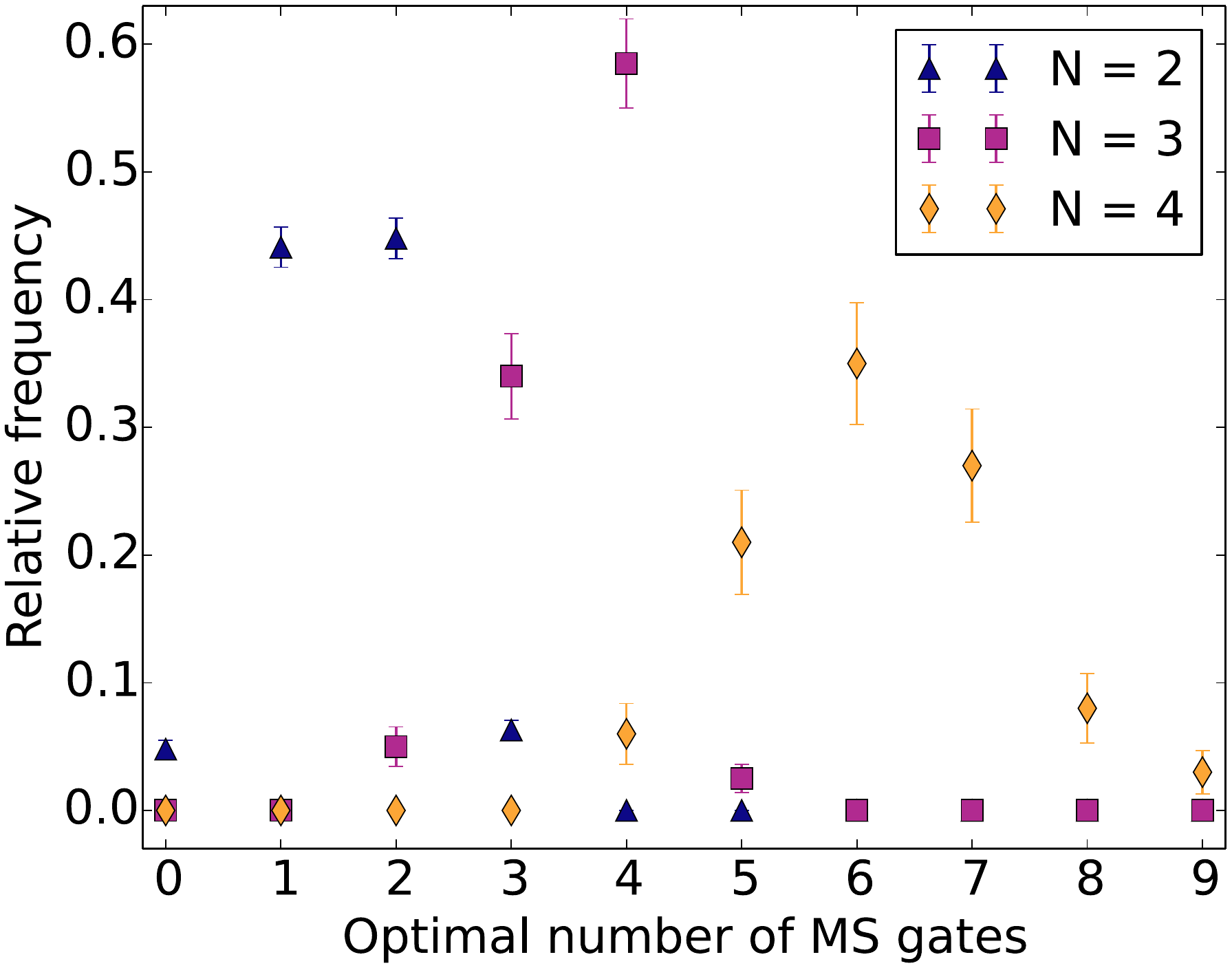}
\caption{Optimal number of entangling gates needed to compile random Clifford operations, for sample sizes $\{1000, 200, 100\}$ for $N = \{2, 3, 4\}$ qubits. Each Clifford gate was generated using $10 N^8$ steps of the random walk described in Ref.~\cite{DiVincenzo2001}. Error bars correspond to one standard deviation.}
\label{fig:statistics}
\end{figure}

A particularly interesting group of unitaries are Clifford gates, which find applications in quantum error correction~\cite{Gottesman1999}, randomized benchmarking~\cite{Knill2008}, and state distillation protocols~\cite{Nielsen2004}. To explore the difficulty of compiling such gates, we have tested our algorithm with randomly generated Clifford gates, as explained in Ref.~\cite{DiVincenzo2001}. We show in Figure~\ref{fig:statistics} the distribution of the optimal number of entangling gates required for compliging two-, three- and four-qubit unitaries. Our results agree with the literature~\cite{Kliuchnikov2013} for the two-qubit case, since MS gates are then equivalent to controlled-Z (or CNOT) gates. For larger numbers of qubits, the performance of our algorithm in terms of number of multi-qubit gates required is also similar to that of algorithms based on two-qubit gates~\cite{Kliuchnikov2013}.

\subsection{Compilation of isometries}

A particular case of interest is the compilation of a unitary whose action only matters on certain input states. This happens, for instance, when one is interested in state preparation starting from some fixed input state. Such operations belong to the more general class of operations known as \emph{isometries}~\cite{Knill2008,Iten2016}. In this case, the problem to be solved has less constraints than when fully specifying the target unitary, so a simpler sequence may be found. In this section we focus on compiling a unitary that is only specified in a particular subspace of the input states, for example:
\begin{equation}
U_{\text{target}} = 
\begin{pmatrix}
u_{11} & u_{12} & \vdots & \vdots\\
u_{21} & u_{22} & \text{free} & \text{free}\\
u_{31} & u_{32} & \vdots & \vdots\\
u_{41} & u_{42} & \vdots & \vdots
\end{pmatrix},
\end{equation}
where the columns marked as `free' are left unspecified. In this case, a suitable fidelity function for the numerical optimization is:
\begin{equation}
  \label{eq:cost-function-restricted}
  f(U) = \left| \operatorname{tr} \left( U|_{S} \, U_{\text{target}}|_{S}^{\dagger} \right) \right|^2,
\end{equation}
where $U|_{S}$ is a rectangular matrix with the components of the unitary in the restricted subspace.

\begin{figure*}[ht]
\centering
\includegraphics[width=.7\linewidth]{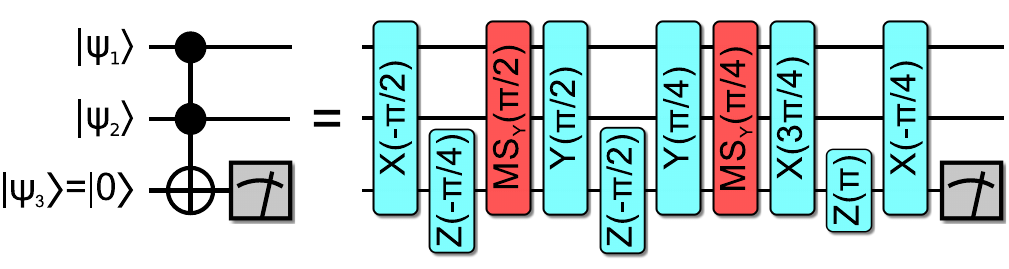}
\caption{Left: a unitary mapping (Toffoli gate) is applied, after which qubit 3 is measured. Right: a pulse sequence for implementing the circuit on the left. This implementation is simpler than in the fully constrained case (figure \ref{fig:toffoli}) because of the additional degrees of freedom when compiling.}
\label{fig:feedback}
\end{figure*}

A more general case is where some of the relative phases of the projections of the unitary acting on different subspaces of the whole Hilbert space are irrelevant. For example, suppose that one wants to apply a unitary to map some observable onto an ancilla qubit and then measure the ancilla, as shown in figure \ref{fig:feedback}. Since the input state of qubit 3 is known to be $\ket{0}$, only the subspace of input states spanned by \{$\ket{000}$, $\ket{010}$, $\ket{100}$, $\ket{110}$\} is relevant. Moreover, the measurement will project the state of the system onto either the subspace spanned by \{$\ket{000}$, $\ket{010}$, $\ket{100}$\}, or that spanned by \{$\ket{111}$\}, and all phase coherence between these alternatives will be lost. Therefore, the compiled sequence can be sought such that it matches the desired unitary in each of the subspaces but allowing an arbitrary phase $\phi$ between them:
\begin{equation}
U_{\text{target}} = 
\begin{pmatrix}
1 & \vdots & 0 & \vdots & 0 & \vdots & 0 & \vdots\\
0 & \vdots & 0 & \vdots & 0 & \vdots & 0 & \vdots\\
0 & \vdots & 1 & \vdots & 0 & \vdots & 0 & \vdots\\
0 & \text{free} & 0 & \text{free} & 0 & \text{free} & 0 & \text{free}\\
0 & \vdots & 0 & \vdots & 1 & \vdots & 0 & \vdots\\
0 & \vdots & 0 & \vdots & 0 & \vdots & 0 & \vdots\\
0 & \vdots & 0 & \vdots & 0 & \vdots & 0 & \vdots\\
0 & \vdots & 0 & \vdots & 0 & \vdots & e^{i \phi} & \vdots\\
\end{pmatrix}
\end{equation}
In this case (figure \ref{fig:feedback}) it is possible to find a simpler implementation than in the fully constrained case (figure \ref{fig:toffoli}), owing to the additional degrees of freedom available, namely arbitrary outputs for the $\ket{\psi_3} = \ket{1}$ input states and an arbitrary relative phase between the two possible measurement outcomes.

In the general case considered here we want to maximize the fidelity in each subspace, without regard to the relative phases between these. Therefore we can seek to maximize the function $f$ consisting of the sum of the fidelity functions (\ref{eq:cost-function-restricted}) corresponding to each subspace:
\begin{equation}
  \label{eq:cost-function-phases}
  f(U) = \sum_j \left| \operatorname{tr} \left( U|_{S_j} \, U_{\text{target}}|_{S_j}^{\dagger} \right) \right|^2,
\end{equation}
where the sum goes over all the subspaces with different relative phases, and $U|_{S_j}$ is a rectangular matrix with the components of the unitary in the $j$-th subspace.

\subsection{Compensation of systematic errors}

Owing to systematic errors, the operations experimentally applied may still be unitary but deviate from the intended ones. An example of this is addressing crosstalk due to laser light leaking onto adjacent qubits. If it is possible to characterize the actual experimental operations being applied, then they can be taken into account for the compilation by adapting our optimization procedure:
\begin{enumerate}
\item Compile the target unitary in terms of the ideal gates.
\item Replace the ideal gates by the experimentally characterized operations.
\item Add operations to obtain a higher fidelity with the ideal target unitary.
\end{enumerate}

\begin{figure*}[htbp]
\centering
\includegraphics[width=.7\linewidth]{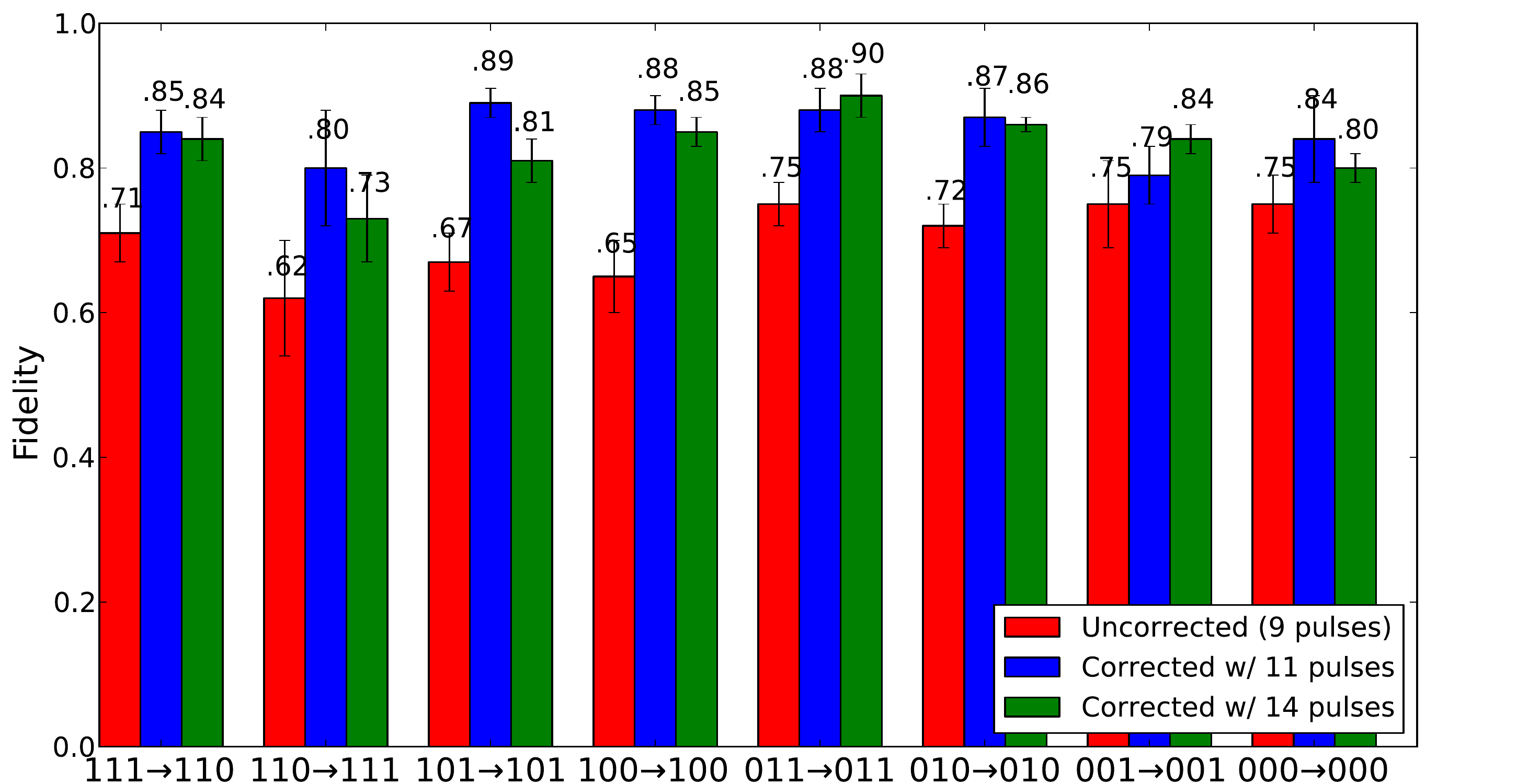}
\caption{State fidelity for a Toffoli gate applied on the 8 canonical input states.}
\label{fig:toffoli-pops}
\end{figure*}

As an example we show that excessive crosstalk can be corrected in an implementation of a Toffoli gate. Figure \ref{fig:toffoli-pops} depicts experimental data corresponding to the action of the Toffoli gate on the 8 input basis states. It can be seen that, by adding just two pulses, the output fidelity for each input state increased in some cases by up to 20\%. The sequence with 11 pulses is actually only an approximate correction to the uncorrected case. The exact correction requires 14 pulses, and actually yields a lower fidelity than the approximate one, since it requires more pulses and each of these has a non-zero error probability.

\section{Conclusions and outlook}

In this work we have shown methods to compile quantum unitaries into a sequence of collective rotations, addressed rotations and global entangling operations. For local unitaries, we have demonstrated an analytic approach that produces the shortest possible sequences in the general case, and adapted the method to simplify the resulting sequences if some constraints on the unitary are lifted. For arbitrary unitaries, we have presented an approach that produces sequences of layered local and entangling operations. This approach is based on a numerical optimization procedure that is faster than previously used ones, and the sequences obtained are by design optimal with respect to the number of entangling gates. Our numerical results suggest upper bounds on the number of $N$-qubit gates required to implement arbitrary three- and four-qubit unitaries.

The results of this paper show that in many cases one may obtain more efficient implementations by considering operations more general than two-qubit entangling gates. However, the exponentially growing complexity of decompositions as the number of qubits increases points to the necessity of keeping the register size small. 

\section{Acknowledgements}

We thank I. Chuang, B. Lanyon, and V. Nebendahl for fruitful discussions. We thank the referees for bringing Refs.~\cite{Knill2008,Iten2016,DiVincenzo2001,Kliuchnikov2013} to our attention. We gratefully acknowledge support by the Austrian Science Fund (FWF), through the SFB FoQuS (FWF Project No. F4002-N16), as well as the Institut für Quantenoptik und Quanteninformation GmbH. E.A.M. is a recipient of a DOC fellowship from the Austrian Academy of Sciences. P.S. was supported by the Austrian Science Foundation (FWF) Erwin Schrödinger Stipendium 3600-N27. This research was funded by the Office of the Director of National Intelligence (ODNI), Intelligence Advanced Reasearch Projects Activity (IARPA), through the Army Research Office grant W911NF-10-1-0284. All statements of fact, opinion or conclusions contained herein are those of the authors and should not be construed as representing the official views or policies of IARPA, the ODNI, or the U.S. Government.

\appendix

\section{Compiling local unitaries}

\subsection{Finding basis changes}
\label{sec:ap-exact}

In this appendix we will show how to satisfy equation (\ref{eq:basis-change}). We need to find a rotation $\coll$ around the equator of the Bloch sphere such that:
\begin{equation}
  \label{eq:basis-change-ap}
  u = \coll^{-1} \sigma_z \coll,
\end{equation}
where $u$ is the generator of a given known unitary $U$, and it can always be written as:
\begin{equation}
  u = \sin \theta \cos \phi \, \sigma_x + \sin \theta \sin \phi \, \sigma_y + \cos \theta \, \sigma_z,
\end{equation}
for some angles $\theta$, $\phi$.

In general $\coll$ is of the form:
\begin{equation}
  \coll = e^{-i \gamma c / 2},
\end{equation}
where $\gamma$ is its rotation angle and $c$ its generator, which must lie on the equator and thus be a linear combination of $\sigma_x$ and $\sigma_y$. If we propose:
\begin{equation}
  c = \sin \phi \, \sigma_x - \cos \phi \, \sigma_y,
\end{equation}
and replace in equation (\ref{eq:basis-change-ap}), we find that the angle of rotation must be:
\begin{equation}
  \gamma = \theta.
\end{equation}

\subsection{Writing a unitary as a product of two equatorial rotations}
\label{sec:ap-two-coll}

We will show here how to decompose an arbitrary unitary as a product of two rotations around the equator of the Bloch sphere, namely:
\begin{equation}
\label{eq:ap-two-coll-decomposition}
U = C_2 C_1.
\end{equation}
The target unitary can be written as:
\begin{align}
U &= \cos \left( \frac{\beta}{2} \right) \id - i \sin \left( \frac{\beta}{2} \right) \times \nonumber \\
& \quad \ \times (\sin \theta \cos \phi \ \sigma_x + \sin \theta \sin \phi \ \sigma_y + \cos \theta \ \sigma_z),
\end{align}
where $\beta$ is its rotation angle,and $\theta, \phi$ determine its rotation axis. Similarly, the equatorial rotations can be written as:
\begin{equation}
C_i = \cos \left( \frac{\alpha_i}{2} \right) \id - i \sin \left( \frac{\alpha_i}{2} \right) (\cos \phi_i' \ \sigma_x + \sin \phi_i' \ \sigma_y ),
\end{equation}
for some rotation angles $\alpha_i$ and phases $\phi_i'$.

We shall asume that:
\begin{align}
\alpha_1 &= \alpha_2 = \alpha,\\
\phi_1' &= \phi + \Delta / 2,\\
\phi_2' &= \phi - \Delta / 2.
\end{align}
Replacing these into (\ref{eq:ap-two-coll-decomposition}) and solving for $\alpha$ and $\Delta$ we obtain:
\begin{align}
\cos^2 \left( \frac{\alpha}{2} \right) &= \frac{1}{2} \left( \cos \left( \frac{\beta}{2} \right) + 1 \right) \sin^2 \theta,\\
\cos \Delta &= \frac{\cos^2 \left( \frac{\alpha}{2} \right) - \cos \left( \frac{\beta}{2} \right)}{1 - \cos^2 \left( \frac{\alpha}{2} \right)}.
\end{align}

\subsection{Unitaries up to a collective Z rotation}
\label{sec:up-to-collective}

Suppose that the unitary $U$ we want to implement is followed by gates whose phase can be freely chosen. Then it must only be specified up to an arbitrary collective rotation $Z'$, since this phase can be absorbed in the following gates. To compile $U$, we shall consider a decomposition of the form (\ref{eq:complete-z}):
\begin{equation}
U = Z' \coll_N Z_{N-1} \coll_{N-1} \dotsm Z_2 \coll_2 Z_1 \coll_1.
\end{equation}
Such a decomposition is more convenient in this case because the last addressed pulse $Z_N$ has been eliminated by taking advantage of the additional degree of freedom provided by $Z'$. We can now follow the same steps as in section \ref{sec:local}. The unitary $\coll_N$ is given by:
\begin{equation}
  \label{eq:coll-n-2}
  \coll_N = Z'^{-1} \, U_N  \coll_1^{-1} \coll_2^{-1} \dotsm \coll_{N-1}^{-1},
\end{equation}
and eliminating this factor from the rest of the equations we obtain:
\begin{align}
  \label{eq:system-eliminated-2}
U_N^{-1} U_1 &= \coll_1^{-1} Z_1 \coll_1,\\
U_N^{-1} U_2 &= \coll_1^{-1} \coll_2^{-1} Z_2 \coll_2 \coll_1,\nonumber\\
&\ \ \vdots \nonumber\\
U_N^{-1} U_{N-1} &= \coll_1^{-1} \coll_2^{-1} \dotsm \coll_{N-1}^{-1} Z_{N-1} \coll_{N-1} \dotsm \coll_2 \coll_1.\nonumber
\end{align}

Equations (\ref{eq:system-eliminated-2}) can be satisfied in exactly the same way as explained in section \ref{sec:local}. In order to satisfy equation (\ref{eq:coll-n-2}) we need to find a rotation $Z'$ such that the generator of $\coll_N$ lies on the equator. This can be done as follows.

We wish to find how to satisfy equation (\ref{eq:coll-n-2}). For this we need to find a rotation $Z$ around the Z axis and a rotation $\coll$ around an axis on the equator of the Bloch sphere such that, for a given unitary $U$, the following equation holds:
\begin{equation}
  \label{eq:basis-change-ap-2}
  \coll = Z U.
\end{equation}
$U$ is in general of the form:
\begin{equation}
  U = e^{-i \alpha u / 2},
\end{equation}
and $Z$ is of the form:
\begin{equation}
  Z = e^{-i \beta \sigma_z / 2}.
\end{equation}

We will first find the angle of rotation $\beta$. If we write out (\ref{eq:basis-change-ap-2}) in terms of the generators of $U$ and $Z$ we have:
\begin{align}
  \coll &= \left( \cos \left( \frac{\beta}{2} \right) \id - i \sin \left( \frac{\beta}{2} \right) \sigma_z \right) \times \nonumber \\
  & \quad \ \times \left( \cos \left( \frac{\alpha}{2} \right) \id - i \sin \left( \frac{\alpha}{2} \right) u \right).
\end{align}
Since the axis of rotation of $\coll$ lies on the equator, its generator must not have any Z component, and thus:
\begin{equation}
  0 = \sin \left( \frac{\beta}{2} \right) \cos \left( \frac{\alpha}{2} \right) + \cos \left( \frac{\beta}{2} \right) \sin \left( \frac{\alpha}{2} \right) u_z,
\end{equation}
that is:
\begin{equation}
  \beta = - 2 \arctan \left( \tan \left( \frac{\alpha}{2} \right) u_z \right).
\end{equation}
Once $\beta$ is known, the unitary on the right-hand side of (\ref{eq:basis-change-ap-2}) is fully determined, and thus $\coll$ as well.

\subsection{Unitaries up to independent Z rotations}
\label{sec:up-to-independent}

Finally, suppose that the unitary we want to implement is defined up to arbitrary independent rotations for each qubit around the Z axis. This is useful if the unitary is followed by a projective measurement, since any final rotation around the measurement axis for any qubit simply adds a phase and will not change the measured probabilities.

Let us again consider a sequence of the form (\ref{eq:complete-z}). The decomposition must now satisfy, for each qubit:
\begin{align}
Z'_1 U_1 &= \coll_N \dotsm \coll_2 Z_1 \coll_1,\\
Z'_2 U_2 &= \coll_N \dotsm  Z_2 \coll_2 \coll_1, \nonumber\\
&\ \ \vdots \nonumber\\
Z'_N U_N &= Z_N \coll_N \cdots \coll_2 \coll_1, \nonumber
\end{align}
where the $Z'_i$ are arbitrary rotations around the Z axis. As before, we can set $Z_N = \id$ and find $\coll_N$:
\begin{equation}
  \label{eq:coll-n-3}
  \coll_N = Z_N' \, U_N  \coll_1^{-1} \coll_2^{-1} \dotsm \coll_{N-1}^{-1}.
\end{equation}
Eliminating $\coll_N$ from the remaining equations we obtain:
\begin{align}
  \label{eq:system-eliminated-3}
U_N^{-1} Z_N'^{-1} Z_1' U_1 &= \coll_1^{-1} Z_1 \coll_1,\\
U_N^{-1} Z_N'^{-1} Z_2' U_2 &= \coll_1^{-1} \coll_2^{-1} Z_2 \coll_2 \coll_1,\nonumber\\
&\ \ \vdots \nonumber\\
U_N^{-1} Z_N'^{-1} Z_{N-1}' U_{N-1} &= \coll_1^{-1} \dotsm \coll_{N-1}^{-1} Z_{N-1} \coll_{N-1} \dotsm \coll_1.\nonumber
\end{align}

Each equation has now an extra degree of freedom coming from the angle of the $Z'_k$ rotation. Let us for simplicity consider the case where the number of qubits $N$ is odd. If we group equations (\ref{eq:system-eliminated-3}) in pairs we get two degrees of freedom per pair, which can be used to remove one of the global operations. Therefore we will discard every even-numbered global operation $\coll_{2k}$ from our decomposition and look for the solution of the following system of equations:
\begin{align}
\label{eq:local-independent-system-pairs}
U_N^{-1} Z_1'' U_1 &= \coll_1^{-1} Z_1 \coll_1,\\
U_N^{-1} Z_2'' U_2 &= \coll_1^{-1} Z_2 \coll_1, \nonumber\\
U_N^{-1} Z_3'' U_3 &= \coll_1^{-1} \coll_3^{-1} Z_3 \coll_3 \coll_1, \nonumber\\
U_N^{-1} Z_4'' U_4 &= \coll_1^{-1} \coll_3^{-1} Z_4  \coll_3 \coll_1, \nonumber\\
&\ \ \vdots \nonumber \\
U_N^{-1} Z_{N-2}'' U_{N-2} &= \coll_1^{-1} \dotsm \coll_{N-2}^{-1} Z_{N-2} \coll_{N-2} \dotsm \coll_1, \nonumber\\
U_N^{-1} Z_{N-1}'' U_{N-1} &= \coll_1^{-1} \dotsm \coll_{N-2}^{-1} Z_{N-1} \coll_{N-2} \dotsm \coll_1, \nonumber
\end{align}
where $Z_k'' = Z_N'^{-1} Z'_{k}$. If the number of qubits $N$ is even, then the last equation is simply left unpaired. It is easy to verify that for each pair of equations the right-hand sides commute, and therefore we must have:
\begin{equation}
[U_N^{-1} Z_{2k-1}'' U_{2k-1}, U_N^{-1} Z_{2k}'' U_{2k}] = 0,
\end{equation}
or equivalently:
\begin{equation}
  \label{eq:cond-3}
[Z_{2k-1}'' U_{2k-1} U_N^{-1}, Z_{2k}'' U_{2k} U_N^{-1}] = 0.
\end{equation}

In order to solve equation (\ref{eq:cond-3}) we need to find rotations $Z_1 = Z(\beta_1)$, $Z_2 = Z(\beta_2)$ that satisfy a general equation of the form:
\begin{equation}
\label{eq:ap-begin}
  [Z_1 U_1, Z_2 U_2] = 0,
\end{equation}
for given arbitrary $U_1$, $U_2$, whose generators are $u_1$ and $u_2$ respectively.

Let us define:
\begin{equation}
  V_i = Z_i U_i,
\end{equation}
and let $v_i$ be the generators of the $V_i$. In order to satisfy (\ref{eq:ap-begin}), the $v_i$ must satisfy:
\begin{equation}
  v_1 = v_2 = v,
\end{equation}
since if two unitaries commute their generators must be the same. Our first goal is to determine the generator $v$. Let us consider the unitary:
\begin{equation}
  W_i = Z_i^{1/2} U_i Z_i^{1/2}.
\end{equation}
By writing down $W_i$ explicitly in terms of the generators of each factor, it can be seen that its generator $w_i$ satisfies:
\begin{equation}
\label{eq:acom-w}
  \{w_i, [\sigma_z, u_i]\} = 0.
\end{equation}
Since we have:
\begin{equation}
  V_i = Z_i^{1/2} W_i Z_i^{-1/2},
\end{equation}
from equation (\ref{eq:acom-w}) we see that:
\begin{equation}
  \label{eq:ap-perpendicular}
  \left\{ v, Z_i^{1/2} [\sigma_z, u_i] Z_i^{-1/2} \right\} = 0.
\end{equation}
The geometrical meaning of this equation is that the vector defined by $v$ on the Bloch sphere is perpendicular to that defined by $Z_i^{1/2} [\sigma_z, u_i] Z_i^{-1/2}$. Since (\ref{eq:ap-perpendicular}) must hold for $i = 1, 2$, $v$ must correspond to the cross product of these vectors:
\begin{equation}
  v = \mathcal{N} \left[ Z_1^{1/2} [z, u_1] Z_1^{-1/2}, Z_2^{1/2} [\sigma_z, u_2] Z_2^{-1/2} \right],
\end{equation}
where $\mathcal{N}$ is chosen such that:
\begin{equation}
  \frac{1}{2} \operatorname{tr}(v^2) = 1.
\end{equation}

Having found $v$, it remains to find the rotation angles $\beta_i$. Now, $v$ must satisfy $[Z_i U_i, v] = 0$, and therefore:
\begin{equation}
  U_i v U_i^{-1} = Z(\beta_i)^{-1} v Z(\beta_i).
\end{equation}
Both $v$ and $U_i$ are known, so $v$ and $U_i v U_i^{-1}$ can be written down explicitely as: 
\begin{align}
  v &= \sin \theta \cos \phi \, \sigma_x + \sin \theta \sin \phi \, \sigma_y + \cos \theta \, \sigma_z, \\
  U_i v U_i^{-1} &= \sin \theta \cos \phi'_i \, \sigma_x + \sin \theta \sin \phi'_i \, \sigma_y + \cos \theta \, \sigma_z,
\end{align}
and therefore:
\begin{align}
  \beta_i = \phi - \phi'_i.
\end{align}

We have shown how to find suitable rotations $Z''$ that fulfill condition (\ref{eq:cond-3}). Once these are found, all the left-hand sides of (\ref{eq:local-independent-system-pairs}) are known unitaries and the system can be solved as before. The last collective rotation $\coll_N$ can be determined from (\ref{eq:coll-n-3}) as shown in appendix \ref{sec:up-to-collective}. We have thus shown how to compile the sought unitary $U$ into a sequence of the form:
\begin{equation}
U = \begin{cases}
\coll_N Z_{N-1} Z_{N-2} \coll_{N - 2} \dotsm \coll_3 Z_2 Z_1 \coll_1 & \text{for odd $N$,}\\
\coll_N Z_{N-1} \coll_{N-1} \dotsm \coll_3 Z_2 Z_1 \coll_1 & \text{for even $N$}.
\end{cases}
\end{equation}

\bibliography{manuscript}

\end{document}